\documentclass[%
 aip,
 amsmath,amssymb,
 reprint,%
]{revtex4-1}

\usepackage{graphicx}% Include figure files
\usepackage{dcolumn}% Align table columns on decimal point
\usepackage{bm}% bold math
\usepackage[utf8]{inputenc}
\usepackage[T1]{fontenc}
\usepackage{etoolbox}

\usepackage{graphicx}
\usepackage{color}
\usepackage{epsf}
\usepackage{amsmath}
\usepackage{amssymb}
\usepackage[english]{babel}
\usepackage{bm}
\usepackage{graphics}
\usepackage{color}

\usepackage{dcolumn}
\usepackage{sidecap}
\usepackage{floatrow}

\usepackage{algorithm}
\usepackage{algpseudocode}

\newcommand{\bes}{ \begin{equation} \begin{split} }
\newcommand{\ees}{ \end{split} \end{equation} }

\usepackage[T2A]{fontenc}
\usepackage[utf8]{inputenc}
\usepackage[english]{babel}

\newcommand{\ignore}[1]{}

 \newcommand{\red}[1]{#1}

\usepackage{lineno,hyperref}
\usepackage{xcolor}

\makeatletter
\def\@email#1#2{%
 \endgroup
 \patchcmd{\titleblock@produce}
  {\frontmatter@RRAPformat}
  {\frontmatter@RRAPformat{\produce@RRAP{*#1\href{mailto:#2}{#2}}}\frontmatter@RRAPformat}
  {}{}
}%
\makeatother
\begin{document}

\preprint{AIP/123-QED}

\title[Oscillation Quenching and Coupling Functions]{Oscillation Quenching Induced By Time-Varying Coupling Functions}

\author{Dushko Stavrov}  
\affiliation{Faculty of Electrical Engineering and Information Technologies, Ss.\ Cyril and Methodius University, Skopje, North Macedonia}

\author{Aneta Koseska} 
 \affiliation{Cellular Computations and Learning, Max Planck Institute for Neurobiology of Behavior-caesar,
Bonn, Germany}

\author{Tomislav Stankovski}\email{aneta.koseska@mpinb.mpg.de,t.stankovski@lancaster.ac.uk}

\affiliation{Department of Medical Physics, Faculty of Medicine, Ss.\ Cyril and Methodius University, Skopje, North Macedonia}
\affiliation{Department of Physics, Lancaster University, Lancaster, UK\\
 }%

% Force line breaks with \\
% \author{A. Author}
%  \altaffiliation[Also at ]{Physics Department, XYZ University.}%Lines break automatically or can be forced with \\
% \author{B. Author}%
%  \email{Second.Author@institution.edu.}
% \affiliation{ 
% Authors' institution and/or address%\\This line break forced with \textbackslash\textbackslash
% }%

 %\homepage{http://www.Second.institution.edu/~Charlie.Author.}
%\affiliation{%
%Second institution and/or address%\\This line break forced% with \\
%}%

\date{\today}% It is always \today, today,

\begin{abstract}
The oscillatory dynamics of natural and man-made systems can be disrupted by their time-varying interactions, leading to oscillation quenching phenomena in which the oscillations are suppressed. We introduce a framework for analyzing, assessing, and controlling oscillation quenching using coupling functions. Specifically,by observing limit-cycle oscillators we investigate the bifurcations and dynamical transitions induced by time-varying diffusive and periodic coupling functions. We studied the transitions between oscillation quenching states induced by the time-varying form of the coupling function while the coupling strength is kept invariant. The time-varying periodic coupling function allowed us to identify novel, non-trivial inhomogeneous states that have not been reported previously. Furthermore, by using dynamical Bayesian inference we have also developed a Proportional Integral (PI) controller that maintains the oscillations and \red{prevents oscillation quenching from occurring}. In addition to the present implementation and its generalizations,  the framework carries broader implications for identification and control of oscillation quenching in a wide range of systems subjected to time-varying interactions. 
\end{abstract}

\maketitle

\begin{quotation}
Natural or man-made oscillations can undergo profound qualitative transformations, for example, by ceasing to exist, a phenomenon known as oscillation quenching. Such transitions may occur through mechanisms like amplitude or oscillation death, often resulting from interactions among oscillatory systems. These events can have severe and sometimes catastrophic consequences, including the long-term extinction of cyclic species in ecological systems, suppression of oscillations in the ocean, or cardiac arrest during myocardial infarction. Recent advances in the study of coupling functions have provided new insights into the mechanisms governing these interactions, offering powerful tools for analyzing and predicting oscillation quenching phenomena. In this work, we integrate these perspectives by introducing a comprehensive framework for the analysis, assessment, and control of oscillation quenching through the use of time-varying coupling functions, thereby deepening the understanding of quenching mechanisms and their associated transitions.
\end{quotation}

\section{\label{sec:intro}Introduction}

Natural and man-made systems are rarely isolated, instead they interact and influence each other's dynamics through complex coupling functions. Examples include physiological systems such as neuronal, genetic or cardiovascular systems \cite{Ermentrout:90,SuarezVargas:09,Manasova:23,Lehnertz:14,Zarghami:20,Stankovski:12b,Lukarski:22,Stefanovska:00a}, ecological and climate models \cite{Gallego:01,Hastings_2018}, electronic circuits \cite{SuarezVargas:09,Kocarev:95,Ruwisch:99}, lasers \cite{Kim:05} etc. The coupling functions of such systems can change in time and space. The proper functioning of these systems relies on their oscillatory dynamics, and it is imperative that these dynamics are maintained even under time-varying interactions. However, it has been shown that oscillations can cease under particular conditions and oscillation quenching can occur  \cite{Koseska:13}. Oscillation quenching is an interaction phenomenon in which the oscillations are suppressed, where they eventually die off and cease to exist \cite{Koseska:13}. The two most common types of oscillation quenching are oscillation death (OD) and amplitude death (AD). Such quenching can have extreme and catastrophic outcomes for oscillators, including for example a long term extinction of a cyclic species in ecology \cite{Hornfeldt:04,Hastings_2018}, suppression of oscillations in the Atlantic and Pacific oceans in climate \cite{Gallego:01}, and stopping of a human heart during a myocardial infarction \cite{Reed:17,Andersen:19}.

\begin{figure*}
\floatbox[{\capbeside \thisfloatsetup{capbesideposition={right,top},capbesidewidth=5.25cm}}]{figure}[\FBwidth]
{\caption{\textbf{Bifurcation analysis of the oscillation quenching induced by time-varying diffusive coupling function.} (a) the temporal evolution of the coupling coefficients $ c_1(t)$ and $c_2(t) $. (b) and (c) are the state responses $ x_1(t)$ and $x_2(t) $ of the two oscillators, illustrating the transitions from Limit Cycle (LC) to Amplitude Death (AD) and Oscillation Death (OD). The enlarged inset shows the LC oscillations. (d) bi-parameter ($c_1,c_2$) bifurcation diagram, showing the AD, OD, and LC regions. (e)–(f) showing the specific bifurcations, including Saddle-Node (SN), Hopf bifurcation (HB), and Pitchfork bifurcation (PB).}}
{\includegraphics[width=0.675\textwidth,angle=0]{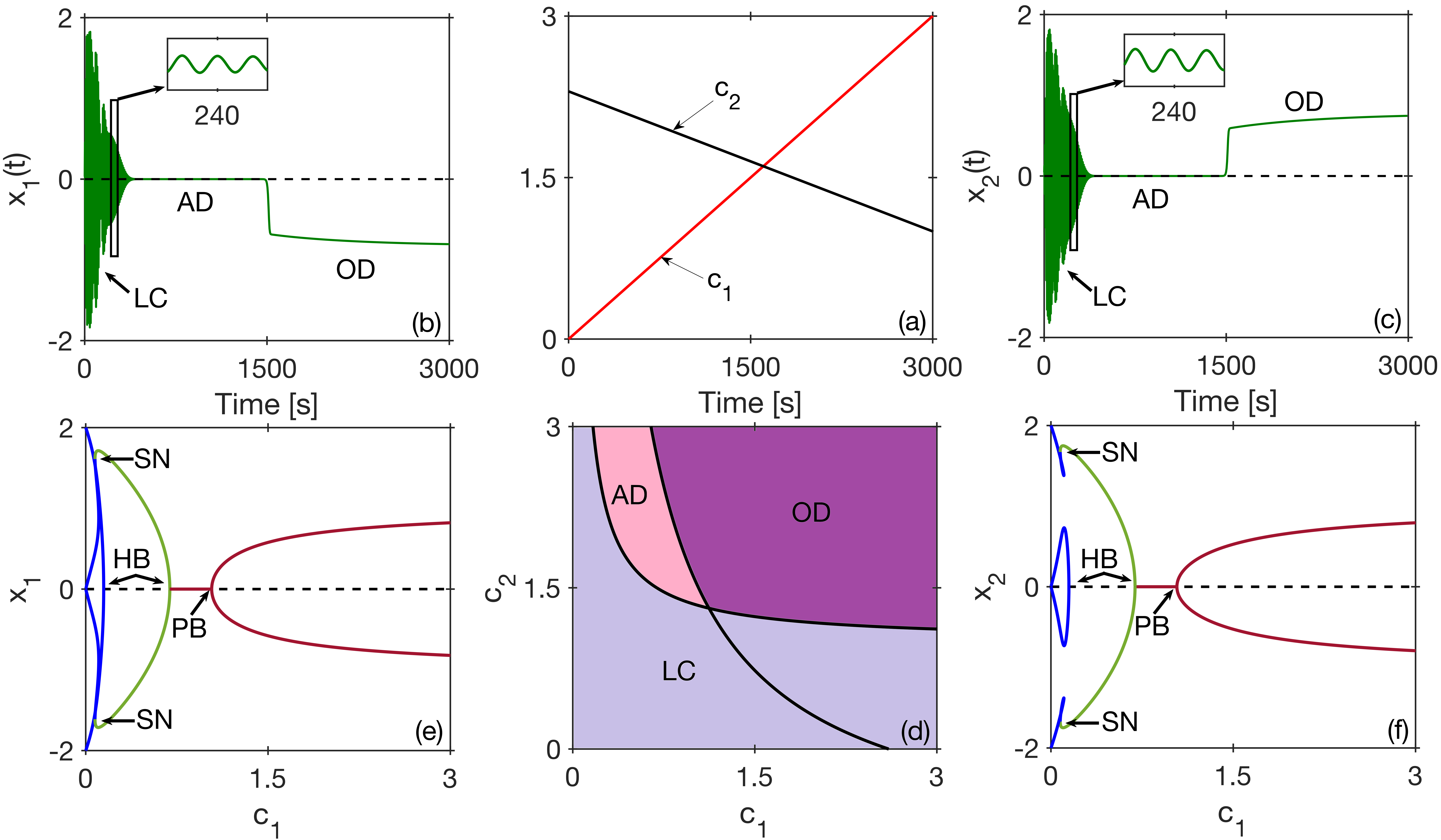}}
 \label{fig:DIFF_VDP}
\end{figure*}

Understanding how oscillation quenching emerges and/or how they can be controlled in coupled systems is therefore of particular practical importance, for which analysis of the underlying \textit{coupling functions} \cite{Stankovski:17b} can be utilized.
The coupling functions describe  \textit{how} the interaction occurs and manifests, thus revealing a functional mechanism. A coupling function has two characteristics: a coupling strength and a form of the coupling function that defines the functional law and the underlying mechanism of the interaction. Oscillation quenching has been shown to emerge for different coupling function forms, such as diffusive coupling \cite{Koseska:13a,Zou:15}, conjugate coupling \cite{Karnatak:07}, mean-field coupling \cite{Sharma:12}, nonlinear coupling \cite{Prasad:10}, environmental coupling \cite{Resmi:12}, and repulsive coupling \cite{Bera:16} etc. However, the effect of time-varying coupling functions on system dynamics has not been investigated so far.

Moreover, the design of powerful methods for the inference and reconstruction of coupling functions\cite{Topal:23,Smelyanskiy:05a,Stankovski:12b,Kralemann:13b,Levnajic:11,Friston:03}  has led to applications in various fields, including chemistry \cite{Kiss:07}, climate \cite{Moon:19},
secure communications \cite{Stankovski:14a}, social sciences \cite{Ranganathan:14},  cardiorespiratory interactions \cite{Kralemann:13b,Stankovski:12b} and neuroscience \cite{Bick:20,Omejc:25,Suzuki:18}.  However, a large fraction of these studies on coupling functions focus on synchronization \cite{Kocarev:96,Pikovsky:01} as one of the fundamental phenomena observed in coupled systems, with significantly less work devoted to understanding the mechanisms under which oscillations are quenched. In particular, the question of which time-varying interactions lead to oscillation death (OD) or amplitude death (AD) and how this can be controlled has not been addressed.

In this study, we introduce a framework for characterizing transitions to oscillation quenching in systems of coupled oscillators subjected to time-varying coupling functions. First, we investigated a simpler diffusive coupling function, after which we studied more complex behaviors with a periodic coupling function. We conducted a detailed bifurcation analysis of the different oscillation quenching mechanisms for a case of invariant coupling strength, but with a time-varying form of the coupling function. Additionally, for the periodic coupling functions we identified a novel OD/AD solution consisting of an infinitely large number of non-trivial inhomogeneous states. Finally, we developed a control system strategy for avoiding oscillation quenching and maintaining active oscillations by combining dynamical Bayesian inference and a Proportional Integral (PI) controller.

\section{\label{sec:OCCFs} Oscillation Quenching with Time-Varying Coupling Functions }

\subsection{\label{sec:system} The system of coupled oscillators}

As an archetypal oscillator, first we used the van der Pol oscillator, which is a non-linear and non-conservative system. \red{To demonstrate that the coupling-induced oscillation quenching mechanism has broader implications and is not tied to specific oscillator structure}, in the Supplementary Material (SM), we also studied the case of coupled Stuart-Landau oscillators.
Thus, here we consider a system of two coupled van der Pol oscillators, each given by:
%\begin{equation}
\begin{align}\label{eq:coupled_VDPs}
\dot{x}_{i} &= y_{i}  + q_i(x_i,x_j,t)\\ \nonumber
\dot{y}_{i} &= \mu (1-x_i^2)y_i - \omega_i^2 x_i,
\end{align}
%\end{equation}
here $ i,j = 1,2; i\neq j  $, and the parameters of the individual oscillators are: $ \mu = 0.35 $, $ \omega_1 = \sqrt{0.71}$, and $\omega_2 = \sqrt{0.97}$, \red{where $\mu$ gave nonlinear limit-cycle oscillations and the oscillator frequencies were chosen to be close in order to differentiate coupling-induced quenching from the effects of large parameter mismatch.} We will consider the coupling functions $q_i(x_i,x_j,t)$ to be either diffusive or periodic, as discussed below.

\subsection{\label{sec:CF_diffusive} The case of diffusive coupling function}

We start our analysis with a commonly studied diffusive coupling function. The diffusive interaction is represented by the difference between the state variables of two oscillating systems \cite{Lopes:23,Koseska:13,Zou:15}. We observed the effects of the time-varying diffusive coupling functions on the interactions and how they mediate transitions between the oscillation quenching states. That is, we analyze in detail how the oscillations are quenched and stop to exist, and when there are oscillation death and amplitude death transitions. Using bifurcation analysis \cite{Ermentrout:02}, the transitions are then described by their associate types of bifurcation on the bifurcation diagrams.

Considering the coupled system of oscillators (\ref{eq:coupled_VDPs}), we introduce the diffusive coupling function \cite{Zou:15} defined as:
\begin{equation}
q_i(x_i,x_j,t) = c_1(t) (x_j(t) - c_2(t) x_i(t)), \quad i,j=1,2, \quad i\neq j,
\label{eq:Diff_CF}
\end{equation}
where the temporal behavior of the coupling coefficients $c_1(t)$ and $c_2(t)$ is assumed to be linear and given as:
\begin{equation}
\begin{aligned}
c_1(t) &= 3 \cdot (t/T), \
c_2(t) &= - (1.3/T)\cdot (t-1.5)+2.3.
\end{aligned}
\label{eq:coupling_coeff_Diff_VDP}
\end{equation}
% and they are varied in the interval $t = 0 \rightarrow T$, where $T$ is the observation time. 

% where the coupling coefficients $c_1(t)$ and $c_2(t)$ are varied in time, over $t \ in [0, T]$ and $T$ is the observation time, according to the following linear relations:

% \red{DULE: napisi go sledniot paragraf spored slikata - i pazi prv pat go diskutirame sega tuka vo trudot, pa odi poleka i vovedi ja slikata i analizite postepeno.. -->} 

The coupling coefficients are varied linearly in the interval $t = 0 \rightarrow T$, where $T = 3000$ is the observation time, as shown in Figure \ref{fig:DIFF_VDP} (a). Specifically, $c_1(t)$ increases linearly from 0 to 3, while $c_2(t)$ decreases linearly from 2.3 to 1. Due to this temporal evolution, the states of the system $x_1(t)$ and $x_2(t)$ undergo qualitative transitions between the limit cycle (LC), amplitude death (AD) and oscillation death (OD), as illustrated in Figure \ref{fig:DIFF_VDP} (b)–(c). At first, the oscillators are entrained in a limit-cycle regime (LC). As time progresses, the oscillations are quenched, and consequently, the system transitions into amplitude death (AD) and later into oscillation death (OD). These transitions were also observed through bifurcation analysis. The bi-parameter bifurcation diagram Figure \ref{fig:DIFF_VDP} (d) defines the boundaries of the dynamic regimes LC, AD and OD in $c_1,c_2 \in [0,3]$ parameter space, while the one-parameter bifurcation diagrams Figure \ref{fig:DIFF_VDP} (e)–(f) for $c_1 \in [0,3] $ with fixed $c_2 = 1.5$ confirm the same sequence of dynamical regimes observed in the state time responses. For small $c_1$, unstable limit cycles exist but stabilize via a saddle-node (SN) bifurcation, giving rise to a unique stable limit cycle. This cycle is subsequently destroyed as the homogeneous steady state stabilizes through an inverse Hopf bifurcation (HB). Finally, the homogeneous steady state loses stability via a pitchfork bifurcation (PB), leading to two stable inhomogeneous steady states.

% The coupling coefficients change in a linear fashion Figure \ref{eq:coupling_coeff_Diff_VDP} (a), such that, while $c_1(t)$ linearly increases in the interval $[0, 3]$, the coefficient $c_2(t)$ linearly decreases in the interval $[2.3, 1]$. As a consequence of this temporal behavior, the states $x_1(t)$ and $x_2(t)$ pass through different qualitative states Figure \ref{eq:coupling_coeff_Diff_VDP} (b) and (c). Initially, the states are confined in an oscillatory state (LC), after which, the states go into oscillation quenched state, first in AD and then in OD. This example shows that the temporal variation of the coupling coefficients and thus the coupling function can definitely be a source of transitions between oscillation and oscillation-quenched states.

\subsection{\label{sec:CF_period} The case of periodic coupling function}

 To observe more complex and non-trivial dynamics, next we studied a periodic coupling function. The motivation to use a periodic coupling function comes from the fact that many systems, processes, or their interactions found in nature manifest periodic variations. This is particularly present in oscillatory interacting systems in biology, such as, for example, the cardiovascular system \cite{Hirsch:81,Stefanovska:99a}. In our previous studies, we have observed and induced a periodic sine time-variation in the cardio-respiratory coupling function \cite{Lukarski:22,Stankovski:12b}. Thus, naturally, we wanted to explore further if, and under what conditions, a periodic coupling function could induce transitions to oscillation quenching.
 We considered an arbitrary coupling function chosen to be periodic and we allowed its form to vary in time:
 
 %The motivation to use the periodic coupling function  (\ref{eq:CF_periodic}) comes from the fact that many systems, processes, or their interactions found in nature manifest periodic variations. This is particularly present in oscillatory interacting systems in biology, like for example in the cardiovascular system \cite{Hirsch:81,Stefanovska:99a}. \blue{To illustrate this further, and following our previous studies \cite{Lukarski:22,Stankovski:12b}, in the SM \cite{NoteCRS} we present an example of a periodic cardio-respiratory time-varying coupling function.} The letter presents an analysis framework which is quite general and can be applied to various models and coupling functions. To demonstrate this, we have provided (in the SM \cite{NoteCRS}) also analysis of coupled Stuart-Landau oscillators with periodic, but also aperiodic and diffusive time-varying coupling functions.

%The periodic coupling function in both directions was chosen to be:
\begin{equation}
q_i(x_i,x_j,t) = c_1^{i}(t) \sin(1.8 x_j )- c_2^{i}(t) \sin(1.8 x_i )
\label{eq:CF_periodic}
\end{equation}
where $c_1^{i}(t)$ and $c_2^{i}(t)$ are the parameters for the sub-coupling components.

\begin{figure}%[!h]
	\center
	\includegraphics[width=\linewidth]{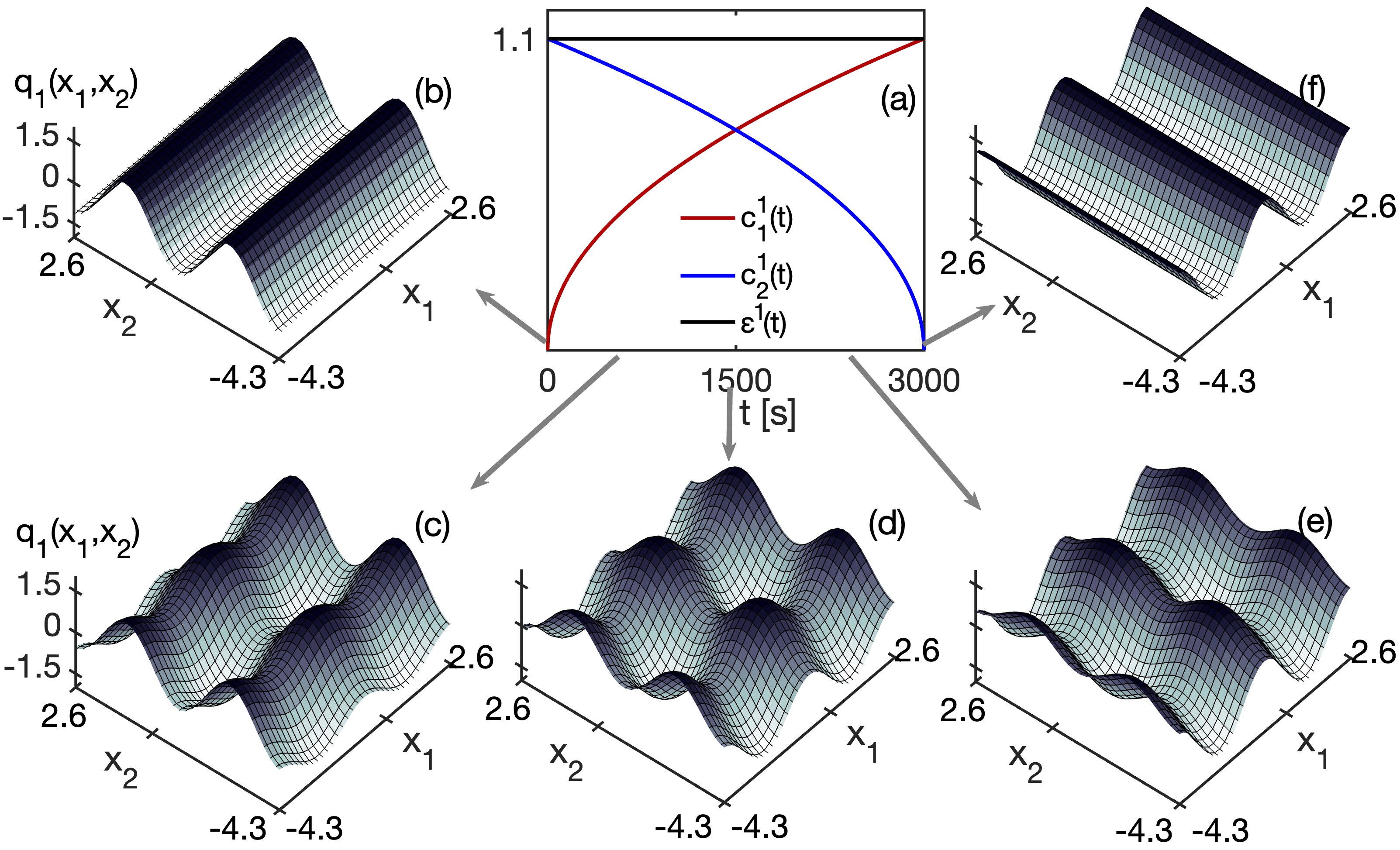}
	\caption{\textbf{Time-varying coupling functions for time-invariant coupling strength.} (a) shows the time-variability of the coupling coefficients $ c^{1}_{1}(t),c^{1}_{2}(t) $and the time-invariant net coupling strength  $ \varepsilon^{1} (t)$. (b)-(f) present the variability of the form of coupling function with time -- the arrows indicate the time instance of the coupling function. For comparison the amplitude on z-axis is the same for all $q_{1}(x_1,x_2)$.}
 \label{fig1_TVCFs}
\end{figure}

When a coupling function is time-varying, both the coupling strength and the form can vary in time, often with different rates or proportions. Whereas the variations of the coupling strength are well known, the functional form is less studied and analyzed. Here, we study a special case where the coupling strength is time-invariant, while the form of the coupling function is time-varying \cite{Hagos:19}. To achieve this specific variability of the coupling functions, we set the sub-coupling components $c_1^{1}(t)$ and $c_2^{1}(t)$ in the direction from the second to first oscillator, to vary in time as:
$c_1^{1}(t) = 1.1 \sqrt{t/T},\text{   }\text{   }\text{   } c_2^{1}(t) =1.1 \sqrt{(T-t)/T}.$

In this way, the net coupling strength $\varepsilon^1(t)$, defined as \red{a} Euclidean norm $\varepsilon^1(t)=\sqrt{\{c_1^{1}(t)\}^2+\{c_2^{1}(t)\}^2}$, is an invariant constant $\varepsilon^1(t)=1.1$ throughout the observation time $T$ -- Fig.\ \ref{fig1_TVCFs} (a). The net coupling strength in the other direction, from the first to second oscillator $\varepsilon^2(t)$, and its sub-coupling components are set to be constant $c_1^{2}(t)=0.3$ and $c_2^{2}(t)=0.9$. Because the sub-coupling components are varying, they will also introduce variations in the form of the coupling functions --  Fig.\ \ref{fig1_TVCFs} (b)-(f). By comparing the form from (b) to (f) one can notice how the form is varying in time e.g.\ the starting time on Fig.\ \ref{fig1_TVCFs} (b) and the ending-time coupling function on Fig.\ \ref{fig1_TVCFs} (f) are in opposite directions. 

\subsubsection{\label{sec:CF_perBif} Bifurcation analysis}

\begin{figure}%[!h]
	\center
	\includegraphics[width=\linewidth]{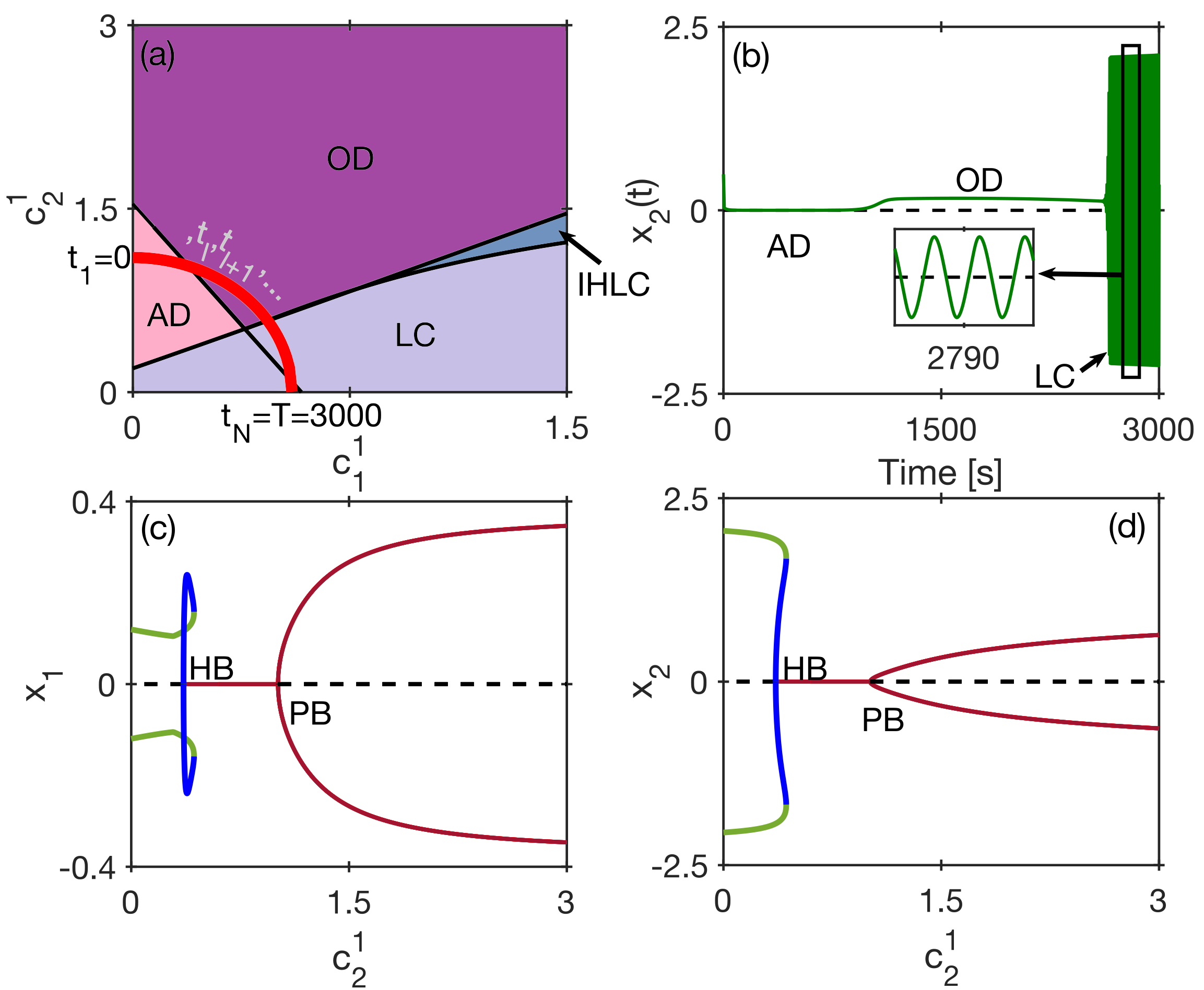}
	\caption{\textbf{Bifurcation analysis of the oscillation quenching induced by time-varying coupling functions.} (a) bi-parametric ($c^{1}_1,c^{1}_2$) bifurcation diagram, showing the Amplitude Death (AD), Oscillation Death (OD), Limit Cycle (LC) and the Inhomogeneous LC (IHLC) regions. The quarter circle (thick red line) indicates the time-variability of $c^{1}_1(t),c^{1}_2(t)$ along the parameter space.  (b) an example of the dynamical time-evolution of $x_2(t)$ showing transition between AD, OD and LC. The enlarged inset shows the LC oscillations.
The parameters $c^{1}_1(t),c^{1}_2(t)$ which induce the transition in (b) are varied along the quarter-circle line in the left-bottom corner of (a). Specific Hopf Bifurcation (HB) and Supercritical Pitchfork Bifurcation (PB) for the $x_1$ state (c) and the $x_2$ state (d) in respect of the $c^{1}_2$ coupling parameter.}
\label{fig2_Bif}
\end{figure}

Next, we introduce these time-varying coupling functions in the system of coupled van der Pol oscillators (Eq. \ref{eq:coupled_VDPs}) and investigate the associated dynamical transitions by means of bifurcation analysis \cite{Ermentrout:02}. 
The biparametric bifurcation diagram in Fig.\ \ref{fig2_Bif} (a), calculated in terms of the two sub-coupling components, shows that there are four characteristic regions: amplitude death (AD), oscillation death (OD), limit cycle (LC) and the inhomogeneous LC (IHLC) region. 
The specific time-variability of the coupling function is presented here as a quarter-circle line in the left-bottom corner of Fig.\ \ref{fig2_Bif} (a). The radius of this quarter-circle is equal to the net coupling strength $\varepsilon^1(t)=1.1$. Therefore, by following the variability of the form of the function on the quarter-circle one can observe the possible transitions between the three regions -- amplitude death, oscillation death and limit cycle. Fig.\ \ref{fig2_Bif} (b) shows the evolution of the state variable $x_2(t)$, where due to the time-varying coupling function the oscillators go through amplitude death, to oscillation death and finally to limit cycle oscillations. The specific types of bifurcations with respect to one of the parameters $c_2^1(t)$ for $x_1(t)$, shown in  Fig.\ \ref{fig2_Bif} (c), shows existence of limit-cycle oscillation which transitions to AD via an inverse Hopf bifurcation, which in turn transitions to oscillation death through Pitchfork bifurcation. Similar results are obtained for the other state variable $x_2(t)$ -- Fig.\ \ref{fig2_Bif} (d).  

\subsection{\label{sec:CF_perNonTrv}  Non-trivial inhomogenous states}

Moreover, for different initial conditions, the system not only exhibits nontrivial steady-states, like Non-trivial Homogeneous Steady State (NHSS) and Inhomogeneous Steady State (IHSS) in their original and known form, but also shows that their number is countless.
Namely, analytical and numerical analysis of the steady states revealed that the coupled systems exhibit infinitely many nontrivial steady-states previously not described, as a result of the periodic coupling function $ q_i $ (Eq.\ \eqref{eq:CF_periodic}), whose values depend on the initial conditions and coupling coefficients. 
Under the mediation of the periodic coupling function, infinitely many nontrivial steady-states, both stable and unstable, are realized and manifest with three distinct states and own basins of attraction. The large number of steady-state positions significantly increases the sensitivity of the coupled system to variations in the initial conditions, external perturbations or noise. In other words, the coupled system can easily undergo transitions between different stable oscillatory or non-oscillatory states.

Without loss of generality, we carried out a semi-analytical analysis of the mode of the infinitely many non-trivial steady states on the system \eqref{eq:coupled_VDPs}, \eqref{eq:CF_periodic} assuming the following values for its parameters: $ \mu = 0.35, \omega_1=\omega_2 = \omega=\sqrt{0.71},  c_1^1(t)=c_1^2(t) = c_1(t), c_2^1(t)=c_2^2(t) = c_2(t) $.  Besides the trivial homogeneous steady state (0,0,0,0), the coupled system possesses infinitely many non-trivial steady states $ (x_1,y_1,x_2,y_2) \neq (0,0,0,0) $, which manifest with three distinct state modes, namely: (i) Non-trivial homogeneous steady states (NHSS) $(x,y,x,y)$, (ii) Inhomogeneous steady states of type 1 (IHSS1) $(x,y,-x,-y)$ and (iii) Inhomogeneous steady states of type 2 (IHSS2) $(x_1,y_1,x_2,y_2)$ where each oscillator state populates different value.
% \\
% 	\noindent a) {Non-trivial homogeneous steady states (\textit{NHSS}) $(x,y,x,y)$}: the states of the individual oscillators $ x_1,x_2 $ and $ y_1,y_2 $ populate equal amplitudes: $ x_1 = x_2=x, y_1=y_2=y$, 
% \\	
% 	\noindent b) {Inhomogeneous steady states of type 1  ({\textit{IHSS1}})  $(x,y,-x,-y)$}: the states of the individual oscillators populate equal amplitudes but opposite in sign, $ \mid x_1 \mid = \mid  x_2 \mid = \mid x \mid , \mid y_1\mid =\mid y_2\mid =\mid y \mid $, and
% \\
% 	\noindent c) {Inhomogeneous steady states of type 2 (\textit{IHSS2})  $(x_1,y_1,x_2,y_2)$}: each state of the individual oscillators populates a distinct value $x_1 \neq y_1 \neq x_2 \neq  y_2$.

The steady states of the system \eqref{eq:coupled_VDPs} are determined by solving the homogeneous system of nonlinear equations obtained by equating the first derivatives of the state variables to zero $ (\dot{x}_{1}(t),\dot{y}_{1}(t),\dot{x}_{2}(t),\dot{y}_{2}(t)) = (0,0,0,0) $, while looking at the stationary solution. If we assume the coupling coefficients $c_1$ and $c_2$ are fixed, then the problem of determining the steady states reduces to solving a system of 4 equations for $(x_1, y_1, x_2, y_2)$. This system has infinitely many solutions (steady states) as a result of the periodicity of the sub-components of the coupling function. From here, it follows that for every distinct realization of the coupling coefficients $c_1$ and $c_2$, there exist infinitely many combinations $(x_1, y_1, x_2, y_2) \neq \textbf{0}$ that represent solutions of the system \eqref{eq:coupled_VDPs}.

By using fixed values of $c_1$ and $c_2$ and solving $\dot{x}_{1}(t)=0$ and  $\dot{x}_{2}(t)=0$ of system \eqref{eq:coupled_VDPs} with respect to $y_1$ and $y_2$, we obtain: 
\begin{equation}
\begin{split} y_1 = -q_1 =  c_2 \sin(1.8 x_1) - c_1 \sin(1.8 x_2),
\\
y_2 = -q_2 = c_2 \sin(1.8 x_2) - c_1 \sin(1.8 x_1).
\nonumber
\end{split}
\end{equation}
Next, if the obtained $ y_1$ and $y_2 $ are substituted in the other equations $\dot{y}_{1}(t)=0$ and  $\dot{y}_{2}(t)=0$, respectively, then the system \eqref{eq:coupled_VDPs} is reduced to two nonlinear equations with two unknowns $x_1$ and $x_2$:
\begin{equation}
\begin{split} 
f_1 &= 0 = \mu (1-x_1^2)\cdot [c_2 \sin(1.8 x_1) - c_1 \sin(1.8 x_2)] - \omega^2 x_1,
\\
f_2 &= 0 = \mu (1-x_2^2) \cdot [c_2 \sin(1.8 x_2) - c_1 \sin(1.8 x_1)] - \omega^2 x_2.
\end{split}
\label{eq:coupled_identical_VDPs_SteadyStates_4variables}
\end{equation}
Based on the system of equations \eqref{eq:coupled_identical_VDPs_SteadyStates_4variables}, an individual analysis of the three types of nontrivial steady states (NHSS, IHSS1 and IHSS2) and the conditions under which they manifest, can be performed.

Let us first examine the conditions for emergence of the \textit{NHSS} steady states. Introducing the substitution: $ x_1 = x_2=x, y_1=y_2=y$, reduces the equations \eqref{eq:coupled_VDPs} to two unique equations, $ y_1 = y_2 = y$ and $ f_1 = f_2 = f$, where the system \eqref{eq:coupled_identical_VDPs_SteadyStates_4variables} obtains the form:

\begin{equation}
f (x,c_1,c_2)= 0 = \mu (1-x^2)[(c_2 - c_1) \sin(1.8 x)] - \omega^2 x.
\label{eq:NHSS_eq}
\end{equation}
The infinitely many solutions $(x,y)$ of equ. \eqref{eq:NHSS_eq} can be determined only numerically since it is a transcendental equation. Hence, if the trivial solution is disregarded, and the following conditions are met:  $ c_1 \neq c_2 $, $ x \neq \pm 1 $, and $ \sin(1.8x)\neq 0 \rightarrow x \neq k\pi/1.8, k \in \mathbb{Z} \setminus \{0\} $, then values for $x$ exist such that equ. \eqref{eq:NHSS_eq} is satisfied. From here, the $ y $ value is determined by substituting for $ x $ in $ y=(c_2 - c_1) \sin(1.8 x) $.

\begin{figure}%[!h]
	\center
	\includegraphics[width=\linewidth]{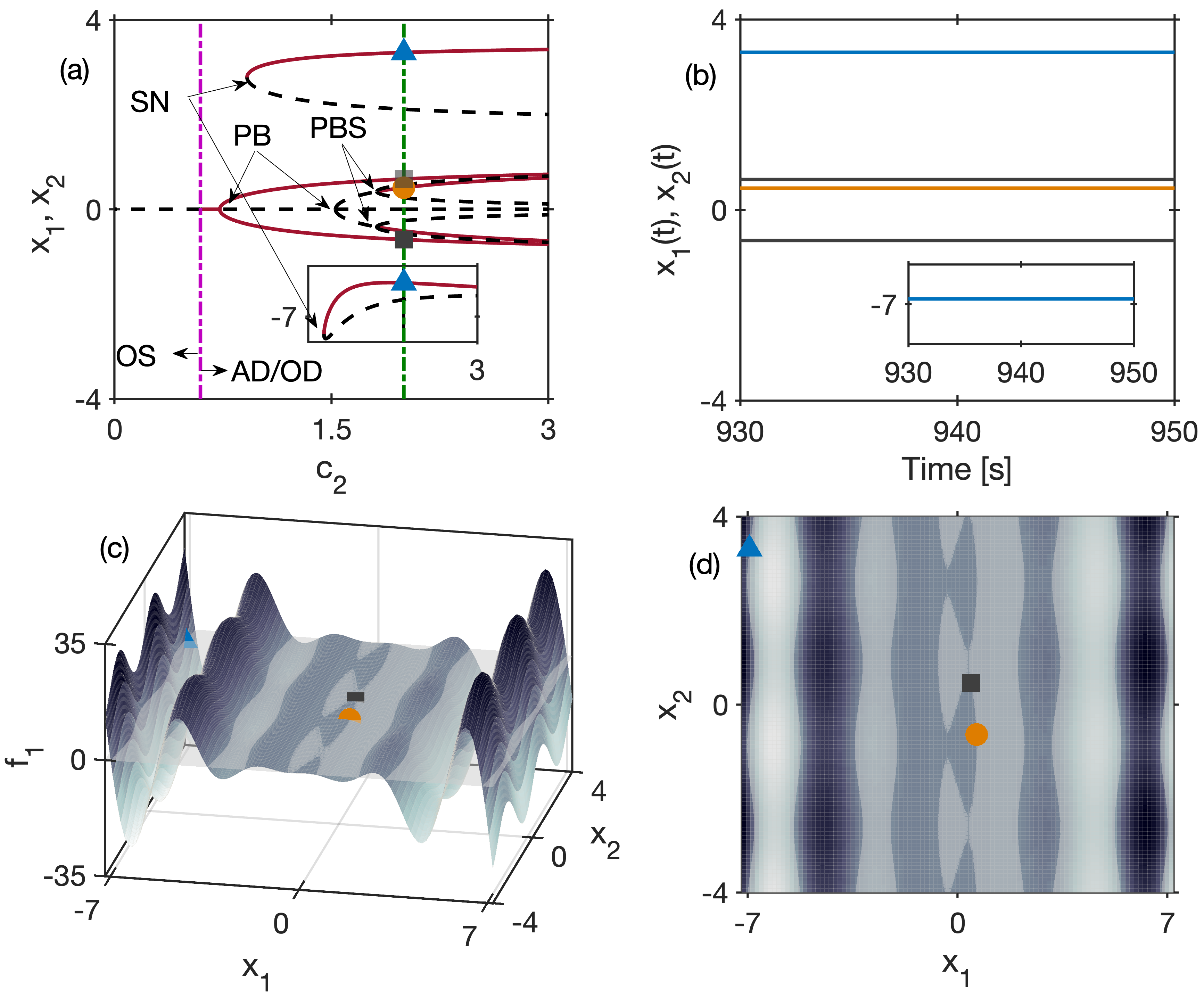}
	\caption{\textbf{Analysis of the periodic OD solutions.} Bifurcation diagram (a) showing the bifurcations including the Saddle Node (SN), PB and Subcritical Pitchfork Bifurcation (PBS). The left vertical dashed line delimits the space into oscillating (OS) and non-oscillating (AD/OD) regions. For better presentation the next bifurcation branch around $x_{1,2}=-7$ is shown on the inset plot. (b) shows the dynamical time-evolution of $x_1(t)$ and $x_2(t)$ on the periodic OD steady states. The solution function $f_1$ of Eq.\ (4) is shown in 3D (c) and 2D (d). The coupling function form dominates the specific solution function $f_1$. Note and compare the markers on (a), (c) and (d) which show the same solution-points for the different OD steady states, as they are shown with same color in (b).  }
 \label{fig:fig3}
\end{figure}

Next, to analyze the conditions for emergence of \textit{IHSS1} steady states of type 1, we introduce the following substitution $(x_1,y_1,x_2,y_2)=(x,y,-x,-y)$. By adopting the same reasoning as above, system \eqref{eq:coupled_VDPs} takes the form:
    \begin{equation}
    f (x,c_1,c_2)= 0 = \mu (1-x^2)[(c_2 + c_1) \sin(1.8 x)] - \omega^2 x.
    \label{eq:IHSS_eq}
    \end{equation}
If we disregard the trivial solution, equ. \eqref{eq:IHSS_eq} will be satisfied only if the following three conditions are met: $ c_1 = c_2 \neq 0, $ $ x \neq \pm 1 $, and $ \sin(1.8x)\neq 0 \rightarrow x \neq k\pi/1.8, k \in \mathbb{Z} \setminus \{0\} $. Once $ x $ solution is found, then $ y $ is determined by substitution in $ y=(c_2 + c_1) \sin(1.8 x) $.

Finally, the \textit{IHSS2} steady states of type 2 emerge for a unique combination of values $ (x_1,y_1,x_2,y_2) \neq \textbf{0} $ which simultaneously reduce the system \eqref{eq:coupled_VDPs} to 0, i.e. $ f_1 = f_2  = 0 $. One can easily show that if $( x_1, y_1,x_2,y_2) $ is a steady state, then $ (x_2,y_2,x_1,y_1)$, $(-x_1,-y_1,-x_2,-y_2)$ and $(-x_2,-y_2,-x_1,-y_1)$ are also steady states.
Further details of the analysis can be found in the Supplementary Material.

The distinct OD solutions are illustrated in Fig. \ref{fig:fig3}. Fig. \ref{fig:fig3} (a) shows the bifurcation diagram for $c_1 = 0.4$ and $c_2 \in [0,3]$ of the system \eqref{eq:coupled_VDPs}. Along the dashed green line, for $c_2=2$, three specific steady states of different type are obtained and demonstrated with markers of different shapes. \textit{NHSS} steady state given by circle marker emerges in the system through Subcritical Pitchfork Bifurcation (PBS), whereas \textit{IHSS1} and \textit{IHSS2} given by rectangle and triangle markers emerge through Supercritical Pitchfork Bifurcation (PB) and Saddle Node Bifurcation (SN), respectively. Their temporal evolution (Fig. \ref{fig:fig3} (b)) is obtained by initiating the system in the vicinity of the steady states. Moreover, to show the dominating role of coupling function in the creation of infinitely many steady states, Fig. \ref{fig:fig3} (c) depicts the intersection between the function $f$, Eq. \eqref{eq:coupled_identical_VDPs_SteadyStates_4variables} and the zero plane in 3D, depicted via the contours, being more prominent in the 2D projection shown in Fig. \ref{fig:fig3} (d). The infinitely many steady states are located along the edges of these contours, as highlighted by the shapes of the three steady states in Fig. \ref{fig:fig3} (c) and even more evident in Fig. \ref{fig:fig3} (d). This shows that the new, and rather complex regime of infinitely many OD solutions can be completely described by the coupling functions.   

Finally, it is worth noting that such a mode of infinitely many nontrivial steady-states can be achieved also with some aperiodic coupling functions. Namely, in our previous analysis the fact that we got infinitely many states was directly linked to the periodicity of the coupling functions. However, one can achieve such infinitely many states also with some other aperiodic unbounded functions. For example, this can be achieved with the use of the following aperiodic coupling function:  

\begin{equation}
q_i (x_i,x_j,c)= c x_i x_j\sin(x_i),
\label{eq:aperiodic}
\end{equation}
where $i\neq j;i,j=1,2.$ Therefore, even though this function is not periodic, for some values of the parameter $c$ it can induce infinitely many nontrivial steady-states.

\section{\label{sec:control}  System control of oscillation quenching with time-varying coupling functions}

Finally, we turn our focus on actively controlling the interactions associated with oscillation quenching phenomena. Namely, we explore the possibility to control (and avoid) the AD/OD regimes using coupling functions, by designing a Proportional Integral (PI) controller \cite{Goodwin:00}.    
The employed PI controller calculates the control term $ \Delta c_1 $ to adjust the value of $ c_1 $ thereby prevent oscillation quenching in the form of amplitude death. The introduction of the control term demands modification in Eq. \ref{eq:coupled_VDPs} of $x_i$ state equations, while the $y_i$ equations are unchanged. The modified $x_i$ state equations are defined as:
\begin{equation}
\begin{matrix}
\dot{x}_{i}(t) = y_{i} (t) + q_i + \Delta c_1(t) sin(1.8 x_j(t))
% \\
% \dot{y}_{i}(t) = \mu (1-x_i^2(t))y_i(t) - \omega_i^2 x_i(t)
\end{matrix},
\label{eq:coupled_VDPs_PI_control}
\end{equation}
and the aggregate value of $c_1$ after the modification is $c_{1,agg} = c_1 + \Delta c_1$,and we used the periodic coupling function $q_i$ Eq.\ \ref{eq:CF_periodic}. Next, the discrete-time PI controller is given as:
\begin{equation}
\Delta c_1 (k) = K_P c_{1,diff} (k) + K_I \sum_{ {m=0}}^k c_{1,diff} (m) N_h h.
\label{eq:PI_controller}
\end{equation}
The $ \Delta c_1$ value is calculated based on the current and previous errors. The error term $ c_{1,diff} $ is calculated as $c_{1,diff} = c_{1,ref}-c_{1,inferred}$ where $ c_{1,ref} $ defines a referent trajectory of $ c_1 $ and $c_{1,inferred}$ is the inferred value of $c_1$ obtained by the Dynamical Bayesian Inference (DBI) method, based on a window of $t_w = 50$ seconds of time-series data \cite{Stankovski:12b}. Further details on the PI control algorithm and the DBI method are given in Supplementary Material. {Because the DBI uses sequential windows in which the evaluation is updated with respect to the prior probabilities, the whole procedure is relatively fast and could be implemented experimentally \cite{Nadzinski:18}.} The tuning parameters of the PI controller are $ K_P = 0.5$, $ K_I = 0.045$, the window length (over which a $\Delta c_1$ value is calculated) is $N_h = 50$ steps and step-size $h = 0.01$ [s]. The coupling parameters vary as: 
\begin{equation}
\begin{split} c_{1,ref}(t) =  0.1 sin((2 \pi/T)t) + 0.8,
\\
c_{1}(t) =  0.5 sin((2 \pi/T)t) + 0.7,
% \nonumber
\end{split}
\end{equation}
and $ c_{2} =  0.6$ is kept constant.

\begin{figure}%[!h]
	\center
 	\includegraphics[width=\linewidth]{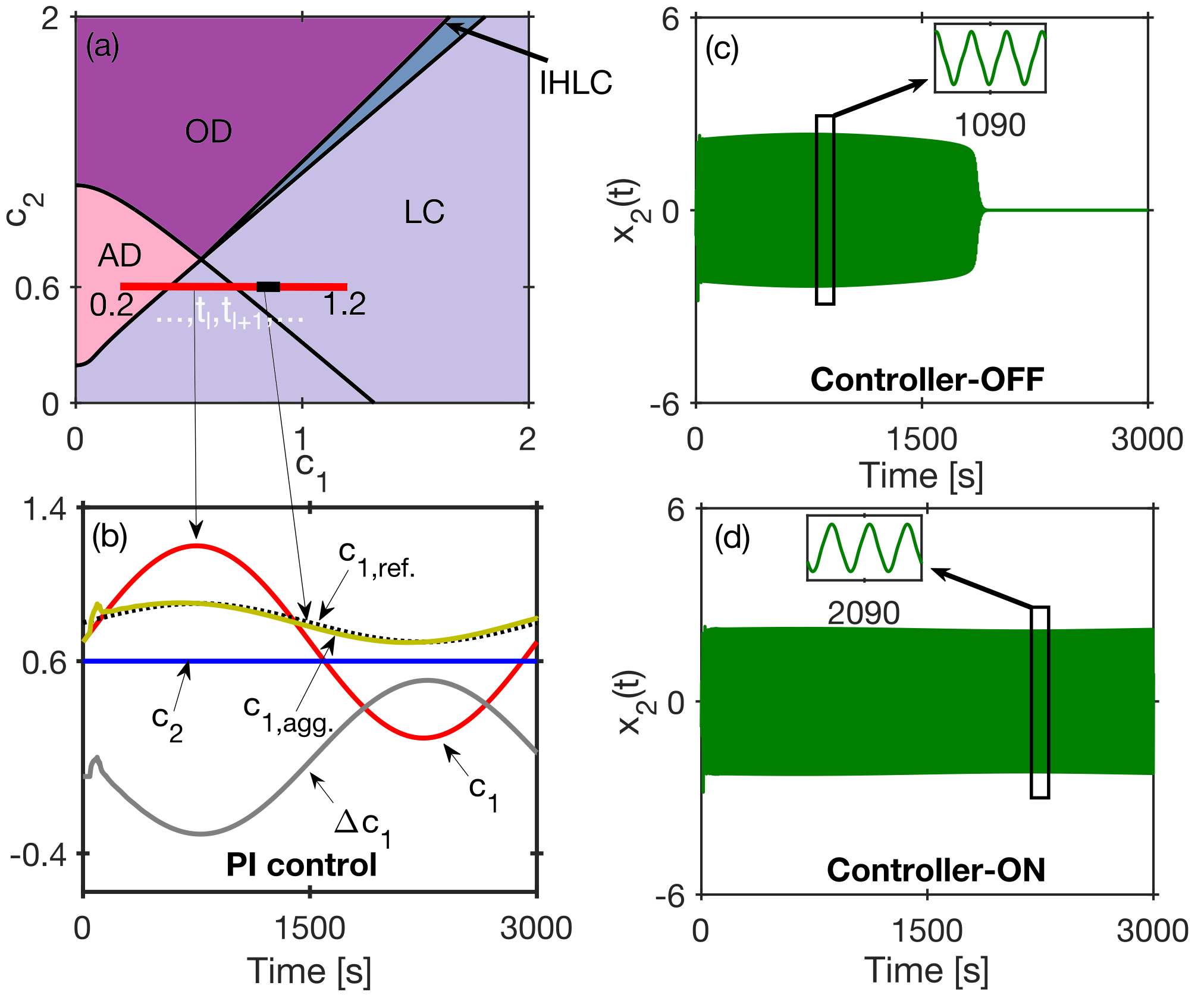}
	\caption{\textbf{Application of a PI system control for avoiding quenching and maintaining active oscillations.} (a) bi-parametric ($c_1,c_2$) bifurcation diagram, showing the AD, OD, LC and IHLC regions. The horizontal line indicates how the coupling parameter $c_1$ varies from LC into AD, while the black part of the line is where the PI controls the variations to avoid AD. (b) shows the evolution of the coupling parameters $c_1$ and $c_2$ and the control parameters: $c_{1,ref}$, $c_{1,agg}$ and $\Delta c_1$. Note the correspondence of $c_1$ and $c_{1,ref}$  between (a) and (b), as indicated with the two arrows. Compare $x_2$ evolution in (c) when there is no control  with (d) when the controller is on preventing  quenching and maintaining active oscillations. }
 \label{fig4_PI}
\end{figure}

The bi-parametric ($ c_{1}$, $ c_{2}$) bifurcation diagram in Fig.\ \ref{fig4_PI} (a) shows the regions in which the system has the desired limit cycle (LC) oscillations, and where it could transition to the undesirable AD and OD quenching. %The goal is thus to maintain the system in the LC region. 
In the autonomous case, the system will transition from LC into AD region -- Fig.\ \ref{fig4_PI} (c), (horizontal line in Fig.\ \ref{fig4_PI} (a)). To maintain the system in the LC regime, we thus first reconstructed the interacting systems from the numerical data using dynamical Bayesian inference \cite{Stankovski:12b}. The inferred  $c_{1,inferred}$ and reference coupling $c_{1,ref}$ were used to calculate the error coupling $c_{1,diff}$ from which we next constructed the control value $\Delta c_1$ -- Fig.\ \ref{fig4_PI} (b). By applying the $\Delta c_1$ control back to the system, the LC oscillations are maintained, thus preventing quenching -- Fig.\ \ref{fig4_PI} (d).  

\section{\label{sec:conclusion} Discussion and conclusions}

We have demonstrated how coupling functions can be employed to identify parametric regions of oscillation quenching in systems with time-varying interactions. While synchronization phenomena have been extensively studied in relation to interaction mechanisms and coupling functions, our focus here is complementary—we investigate how these mechanisms contribute to quenching phenomena. Although various coupling functions have previously been applied to the study of oscillation death and amplitude death, we systematically analyze and evaluate the role of coupling function mechanisms in driving quenching transitions. Furthermore, as this study was motivated by biological oscillatory interactions in nature, we also examine how the time variability of coupling functions influences and mediates oscillation quenching.    

\red{We illustrated the influence of the time-varying coupling functions on oscillation quenching using the van der Pol and Stuart-Landau oscillators. While our analysis focuses on these archetypal systems, the results imply that the same coupling-induced quenching mechanism can be extended to a broader class of self-sustained oscillators that exhibit limit-cycle behavior.} In terms of the coupling functions, we investigated the well-known diffusive coupling function and the biology-inspired periodic coupling function, as we also discussed the implications of an unbounded aperiodic coupling function. 

The time-variability was motivated by biological systems because in nature it is so often observed that the dynamics and the interactions change over time, leading to all sorts of transitions. A typical example is the suppression of brainwave oscillations \cite{Plotkin:81,Williams:45}, or the extreme case of heart myocardial infarction \cite{Reed:17}.   Our approach involved a bifurcation analysis\cite{Ermentrout:02} of the stationary points for the slow time-varying scales. While, due to the temporal variability of the coupling, the system formally represents a nonautonomous one. Therefore, for a more appropriate theoretical treatment, one could also explore nonautonomous bifurcation theory \cite{Rasmussen:07,Kloeden:11}.

As oscillation quenching is quite an extreme change of the oscillation state, often there is a need to control, prevent death or revive the oscillations. There have been different procedures\cite{Ghosh:15,Zou:11,Zou:15} to control oscillation quenching to different effects and extents. Here, we used the PI controller to prevent oscillation death and to maintain the existence of oscillations. The PI control was used to good effect, however, different procedures for system control can be investigated \cite{Goodwin:00,Stavrov:21}, also controlling different aspects of the systems, other than coupling functions.

Furthermore, the generalization extends to the identification method, where in addition to dynamical Bayesian inference, other Bayesian and machine learning methods could potentially perform better for different systems \cite{Berger:96,Friston:03,Dzeroski:95}. 

In summary, we presented a framework which could describe and control quenching with implications for oscillators with varying interactions of different nature, including in mechanics, ecology, climate and medicine \cite{Kato:24,Hornfeldt:04,Hastings_2018,Gallego:01,Lehnertz:14,Andersen:19}. In this way, for example, one could design a control system for myocardial infarction prevention \cite{Osteresch:19,Manson:92}. Therefore, in addition to its current implementation, this study provides a broad framework for the identification and control of oscillation quenching in a wide range of systems subjected to time-varying interactions.
         
\acknowledgments

Our grateful thanks are due to Andrea Duggento, Deniz Eroglu and Tiago Pereira for valuable discussions. AK was supported by the Lise Meitner Excellence Program of the Max Planck Society. TS acknowledges support from the scientific project MON15-6171/24MKD.

%\bibliography{medfizbib}

%

\end{document}